\documentclass{elsart}
\usepackage[numbers,sort&compress]{natbib}
\begin{document}

\begin{frontmatter}
\title{Termination of the $^{12}$C($^{12}$C,$^{12}$C[3$_{1}^{-}$])$^{12}$C[3$_{1}^{-}$]
band of resonances.}
\author[a]{C.A. Bremner}
\author[a]{S.P.G. Chappell}
\author[a]{W.D.M. Rae}
\author[a]{I. Boztosun}
\author[b]{B. Fulton}
\author[b]{D.L. Watson}
\author[b]{B.J. Greenhalgh}
\author[b]{G.K. Dillon}
\author[b]{R.L. Cowin}
\author[c]{M. Freer}
\author[c]{M.P. Nicoli}
\author[c]{\and S.M. Singer}

\address[a]{Nuclear and Astrophysics Laboratory, University of Oxford, Keble
Rd., Oxford OX1 3RH, U.K.}
\address[b]{Department of Physics, University of York, Heslington, York
YO10 5DD, U.K.}
\address[c]{School of Physics and Space Research, University of
Birmingham, Edgbaston, Birmingham B15 2TT, U.K.}

\begin{abstract}
New experimental data for the $^{12}$C+$^{12}$C reaction have been measured in
the centre-of-mass energy range E$_{c.m.}$= 40 to 60 MeV.  Excitation functions
for a number of single and mutual $^{12}$C inelastic channels have been
measured which include the 0$_{gs}$, 2$_{1}^{+}$, 0$_{2}^{+}$, 3$_{1}^{-}$, and
4$_{1}^{+}$ $^{12}$C states. All of the reactions display largely unstructured
excitation functions over this energy range. The absence of further resonances
in this energy region for the
$^{12}$C($^{12}$C,$^{12}$C[3$_{1}^{-}$])$^{12}$C[3$_{1}^{-}$] reaction confirms
theoretical predictions of the termination of the band of resonances found at
lower centre-of-mass energies in this channel.
\end{abstract}

\end{frontmatter}

Heavy-ion resonance reactions have been studied extensively over the past few
decades.  In particular the $^{12}$C+$^{12}$C reaction has drawn significant
interest.  A large number of resonances have been found in, for example,
inelastic reactions including the population of the 2$_{1}^{+}$, 0$_{2}^{+}$,
3$_{1}^{-}$, and 4$_{1}^{+}$ $^{12}$C states
\cite{Cor77,Cor78,Ful80,Wuo92,Wuo94,Cha95,Cha96,Cha98,LeM97}. One particular
theoretical picture which has been used to provide an understanding of the
resonances is one in which the resonances correspond to a molecular band.
Structure models such as the Alpha Cluster Model (ACM) \cite{Mar86},
Hartree-Fock (HF) calculations \cite{Flo84} and Nilsson-Strutinsky (NS)
calculations \cite{Lea75,Rag81} predict a number of rotational bands whose
energy-spin characteristics may be tested against experimental data. A
particularly prominent set of resonances have been observed in the
$^{12}$C(2$^+$)+$^{12}$C(2$^+$) inelastic channel by Cormier {\em et al.}
\cite{Cor77,Cor78}, the energy-spin systematics of these resonances
\cite{Rae87,Rae87a,Sug85,Tan83} agree with that predicted by both the HF
\cite{Flo84}, ACM \cite{Mar86,Zha95} and cranked NS \cite{Rag81} calculations
for the so called F1 configuration. In the ACM this structure corresponds to a
triaxial arrangement of six alpha-particles, which resembles two touching
equilateral triangles. Such a structure has a large overlap with two oblate
$^{12}$C nuclei with their deformation axes aligned. Spin alignment
measurements \cite{Kon85} of the Cormier resonances are also consistent with
the picture of two touching $^{12}$C nuclei orientated as in the F1
configuration. The HF calculations predict that the F1 band should terminate at
a spin of 20~$\hbar$ and the most recent ACM calculations suggest a terminating
spin of 24~$\hbar$ \cite{Zha95,Cha98}. The cranked NS calculations \cite{Rag81}
do not provide a prediction for the termination of the band. More recent
measurements by Chappell {\em et al.} \cite{Cha98} of the
$^{12}$C(3$^-$)+$^{12}$C(3$^-$) inelastic scattering reaction shows evidence
for the continuation of the band of Cormier resonances up to $E_{c.m.}$=43.0
MeV ($E_x(^{24}$Mg)= 57 MeV) and a spin of 22 $\hbar$, very close to the
predicted termination of the F1 band.

The present paper presents a study of inelastic $^{12}$C+$^{12}$C scattering to
states in which the $^{12}$C nuclei are unbound to $\alpha$-decay (including
the $^{12}$C(3$^-$)+$^{12}$C(3$^-$) final state) in the centre-of-mass energy
region $E_{c.m.}$=40 to 60 MeV extending to higher energies than probed in
earlier studies. These measurements provide the first experimental evidence for
the termination of the band of resonances observed in the
$^{12}$C(3$^-$)+$^{12}$C(3$^-$) channel at a spin of 22~$\hbar$, intermediate
between the predictions of the ACM and HF calculations.

The search for this band termination was performed at the Australian National
University(ANU).  An experiment was conducted using the new Charissa strip
detector array located in the MEGHA chamber \cite{Cow97,Cow99}. The array was
composed of eight 500 $\mu$m, 50$\times$50 mm$^{2}$ Si strip detectors
\cite{Cha97}.  These covered an angular range of $\theta_{lab}$ = 5 to
60$^{\circ}$, and an azimuthal angular range $\Delta\phi \approx$ 100 degrees
each side of the beam axis. Each strip detector was divided into 16
position-sensitive strips, providing very high segmentation, and the
possibility of measuring emission angles and thus momenta with the high
precision required for the reconstruction of the reaction kinematics. $^{12}$C
beams of 50 enA intensity ranging from E$_{beam}$= 80 to 120 MeV were incident
upon a 60~$\mu$gcm$^{-2}$ $^{12}$C foil target, producing a data event rate of
5 kHz. The target thickness was found to increase by 25$\%$ during the run
which was corrected for in the analysis of the reaction yields. The higher
energies were obtained using the linear accelerator in conjunction with the
pelletron tandem 14UD accelerator \cite{Oph74,Her74}. Due to the high beam
energies used and the limited detector thickness, events involving
$\alpha$-particles with an energy in excess of $\sim$31 MeV, experience
punch-through (i.e. the particles did not stop in the silicon detectors) and
thus only a fraction of the $\alpha$-particle energy is deposited in the
detector. A Monte-Carlo simulation of the reaction and detection process
indicated that only 4.8{$\%$} of events at the highest beam energy (120 MeV)
experience punch-through for the mutual $^{12}$C(3$_{1}^{-}$) channel. This
fraction is much higher for the other dominant channels, and particularly for
those involving the $^{12}$C(4$_{1}^{+}$) state. Such processes do contribute
 to the overall background levels observed at the higher
energies. However, the reconstruction methods employed in the analysis were
able to clearly distinguish the punch-through events, suppressing this
contribution to the background.

The experimental trigger requirement was a total strip multiplicity of greater
than 3, this implies that the measurement is only sensitive to inelastic
channels in which one of the $^{12}$C nuclei was excited into an
$\alpha$-unbound state. Since the $\alpha$-decay of $^{12}$C feeds states in
$^8$Be, all of which are unbound to $\alpha$-decay a three-$\alpha$ final state
results. Such a decay process occurs for all but the $^{12}$C ground state and
4.4 MeV (2$^+_1)$ states. Thus, for inelastic channels involving a $^{12}$C
nucleus in one of these bound states and the other $^{12}$C in an unbound state
the final multiplicity will be 4. If, however, both $^{12}$C nuclei are excited
to $\alpha$-decaying states then the final state multiplicity is then 6. The
detection of only 5 of the 6 final state particles is sufficient to fully
reconstruct the reaction kinematics. Further details of the reconstruction
techniques can be found in \cite{Cha96,Cha98,Dav90,Shi86,Fre94,Bre01}. The
$\alpha$-decay process is such that the three $\alpha$-particles are emitted
into a cone with an opening angle which is small compared with the detector
geometry. Thus if three hits are observed on one side of the beam axis there is
a large probability that they arise from the decay of $^{12}$C. It should be
noted that there was no explicit particle identification in these measurements,
however the reaction kinematics may in fact be used to identify uniquely
reaction products and decay channels. For example, the reconstruction of the
excitation energy of the parent $^{12}$C nuclei from the momenta of the three
$\alpha$-particles should identify which state was excited in the reaction
process. Such a spectrum is plotted in Figure 1 for a final state consisting of
6 $\alpha$-particles, where the excitation energy for both $^{12}$C nuclei
produced in the collision are reconstructed. The two dimensional spectrum
reveals the mutual excitations of the $^{12}$C nuclei, and the excitation
energies of the states are indicated on the projections. The dominant states in
these spectra are the 0$^+_2$ (7.6542 MeV), 3$^-_1$ (9.641 MeV), 4$^+_1$
(14.083 MeV) states. It is also possible to reconstruct the decay path for the
$^{12}$C $\alpha$-decay, i.e. via the $^8$Be ground state or $^8$Be 3.04 MeV
(2$^+$) state. Also shown in Figure 1 is the decomposition of the excitation
spectrum between these possible decay branches. As expected, the decay of the
4$^+_1$ state proceeds predominantly via the $^8$Be excited state, whilst the
decay of the 0$^+_2$ state feeds the $^8$Be ground state. By placing two
dimensional gates on the spectrum in Figure 1 it is possible isolate the
various mutual excitations, and thus to extract the energy dependence of the
various reaction yields.

The excitation function for the mutual $^{12}$C(3$_{1}^{-}$) excitation is
shown in Figure 2, with a direct comparison with the previous data of Chappell
{\em et al.} \cite{Cha98}. The data of Chappell {\em et al.} were not
normalised for target thickness nor detector efficiency, but only integrated
beam current. Hence the results presented here have been normalised to the
E$_{c.m.}$= 42 MeV data point of their measurements for a comparison of the
structure. Note also that the two experiments do not cover exactly the same
centre-of-mass angular range. Also the present data do not extend low enough in
energy to clearly observe the decrease in cross section below the $E_{c.m.}$=43
MeV resonance. It is clear that no further resonances are observed in this
extended energy range, but only a smooth attenuation of the reaction yield.
Also shown in Figure 2 are the detection efficiencies calculated with Monte
Carlo simulations. These calculations show that the detection efficiency
increases with increasing centre-of-mass energy, and thus the decrease in the
reaction yield cannot be explained by decreasing acceptances. Thus, we believe
that this result shows the termination of the mutual 3$_{1}^{-}$ band of
resonances and thus the highest spin member of the band is J= 22 $\hbar$ at
($E_x(^{24}$Mg)= 57 MeV).

Figures 3 and 4 show the excitation functions for the various reaction channels
which have been observed in the present measurement. These figures show the
experimental cross sections deduced from the normalization to the integrated
beam current, and from calculations of the detection efficiency evaluated as a
function of the centre-of-mass energy. We note that the cross sections
calculated here are in good agreement with earlier measurements
\cite{Wuo94,Ful80}, and that in all instances the statistical errors are
smaller than the data symbols. In Figures 3a and 3b we observe $^{12}$C single
and mutual excitations respectively, involving the 0$_{2}^{+}$, 3$_{1}^{-}$ and
the 4$_{1}^{+}$ channels.  It is clear that none of the channels involving the
0$_{2}^{+}$, 3$_{1}^{-}$ and 4$_{1}^{+}$ states possess the type of resonant
structure that is present at lower energies in the
$^{12}$C(3$_{1}^{-}$)+$^{12}$C(3$_{1}^{-}$) channel. The
$^{12}$C$_{gs}$+$^{12}$C$^*$ reactions generally show a steady decrease in
strength with increasing energy. The increase in yield as the energy decreases
towards $E_{c.m.}$=44 MeV is consistent with earlier measurements of these
reactions \cite{Cha95d} where a resonance was observed in the
$^{12}$C$_{gs}$+$^{12}$C(3$_{1}^{-}$) and $^{12}$C$_{gs}$+$^{12}$C(0$_{2}^{+}$)
reactions at $E_{c.m.}$=41 MeV. There is perhaps some evidence for a small
enhancement in the cross section in the $^{12}$C$_{gs}$+$^{12}$C(4$_{1}^{+}$)
reaction close to $E_{c.m.}$=44 MeV. Enhancements are also observed in the
$^{12}$C(3$_{1}^{-}$)+$^{12}$C(3$_{1}^{-}$) and
$^{12}$C(4$_{1}^{+}$)+$^{12}$C(4$_{1}^{+}$) mutual excitations, shown in Figure
3b, at the same energy. This is close to the large peak previously observed in
the $^{12}$C(3$_{1}^{-}$)+$^{12}$C(3$_{1}^{-}$)reaction \cite{Cha95d} at
$E_{c.m.}$=43 MeV.

Figures 4a and 4b present the remaining mutual channels which have significant
yield in this energy range. Again although broad structures are present, there
is no evidence that strong resonances have been observed. There is some
indication of enhancements at $E_{c.m.}$=44 MeV in the
$^{12}$C(2$_{1}^{+}$)+$^{12}$C(3$_{1}^{-}$),
$^{12}$C(4$_{1}^{+}$)+$^{12}$C(3$_{1}^{-}$),
$^{12}$C(0$_{2}^{+}$)+$^{12}$C(3$_{1}^{-}$) and
$^{12}$C(0$_{2}^{+}$)+$^{12}$C(4$_{1}^{+}$) reactions, and the
$^{12}$C(3$_{1}^{-}$)+$^{12}$C(4$_{1}^{+}$) channel shows weak evidence for a
further structure at $E_{c.m.}$=46~MeV. But, in general all of the reaction
channels demonstrate a rather smooth energy dependence over the measured energy
range. It is clear that over the centre-of-mass energy range $E_{c.m.}$=40 to
60 MeV that the inelastic reactions involving the population of
$\alpha$-particle unbound states are dominated by those which include the
excitation of the 14.083 MeV (4$^+$) state. Thus, coupled channel calculations
which attempt to reproduce the energy dependence of the inelastic scattering
over this energy region must include the coupling to this state. For example, a
number of coupled channel studies of the energy dependence of the inelastic
channels have been performed \cite{Hir94,Ito01,Boz01} which do not include such
couplings.

We have presented data showing the termination of the mutual
$^{12}$C(3$_{1}^{-}$) band of resonances as predicted by the cranked
Bloch-Brink $\alpha$-cluster model of Marsh and Rae \cite{Mar86}. This would
confirm the predictions that the 2$^+$+2$^+$ and 3$^-$+3$^-$ resonances are
associated with the triaxial F1 configuration in $^{24}$Mg. The the observation
that the band terminates at a spin of 22~$\hbar$, intermediate between the ACM
and HF calculations should allow these models to be further refined. We have
also presented excitation functions for the dominant single and mutual $^{12}$C
reaction channels in this energy range. On the whole these reactions appear to
possess a smooth energy dependence, and lack of distinct structures in the
excitation functions indicates that resonant processes are not dominant over
the energy range $E_{c.m.}$=40 to 60 MeV. These measurements indicate that over
this energy range that reactions involving the population of the 14.083 MeV
(4$^+$) state dominate over other inelastic scattering reactions involving the
population of states above the $^{12}$C $\alpha$-decay threshold.

The authors wish to thank the staff of the Department of Nuclear Physics at the
Australian National University for assistance in running the experiments.  We
acknowledge the financial support of the U.K. Engineering and Physical Sciences
Research Council (EPSRC). The experimental work was performed under a formal
agreement between the EPSRC and ANU.

\newpage
\section*{Figure Captions}
Fig. 1. The reconstructed $^{12}$C excitation energies for six fold
$^{12}$C*$^{12}$C* events at 120 MeV. The one dimensional projections indicate
excitation energies of the states observed in the reaction. The strength of the
decay of each state via either the $^8$Be ground state or first excited state
is indicated by the ancillary lines on the one dimensional spectra.  Also
indicated on the two dimensional spectrum are the gates used to filter the data
for angular correlation and cross-section measurements to be performed
(horizontal and vertical dotted lines).\\
\\
Fig. 2. Experimental excitation functions for the
$^{12}$C($^{12}$C,$^{12}$C[3$_{1}^{-}$])$^{12}$C[3$_{1}^{-}$] channel. Results
are shown for a comparison with the previous data of Chappell \cite{Cha98} with
the present data. Also indicated are the Monte Carlo simulation for the
detection efficiencies for each of the experimental data sets. Note that the
present data have been normalised to the $E_{c.m.}$=43 MeV data point of the
previous measurement \cite{Cha98}.\\
\\
\\
Fig. 3. Experimental excitation function for the (a)
$^{12}$C($^{12}$C,$^{12}$C[0$_{2}^{+}$,3$_{1}^{-}$,4$_{1}^{+}$])$^{12}$C(g.s.)
reactions, and (b) for the mutual $^{12}$C(0$_{2}^{+}$, 3$_{1}^{-}$, and
4$_{1}^{+}$) reactions.\\
\\
Fig. 4. Experimental excitation function for the observed channels in the
$^{12}$C(2$_{1}^{+}$)+$^{12}$C* reactions, and (b) the
$^{12}$C(0$_{2}^{+}$)+$^{12}$C(3$_{1}^{-}$),
$^{12}$C(0$_{2}^{+}$)+$^{12}$C(4$_{1}^{+}$)and
$^{12}$C(3$_{1}^{-}$)+$^{12}$C(4$_{1}^{+}$) reactions.\\
\\

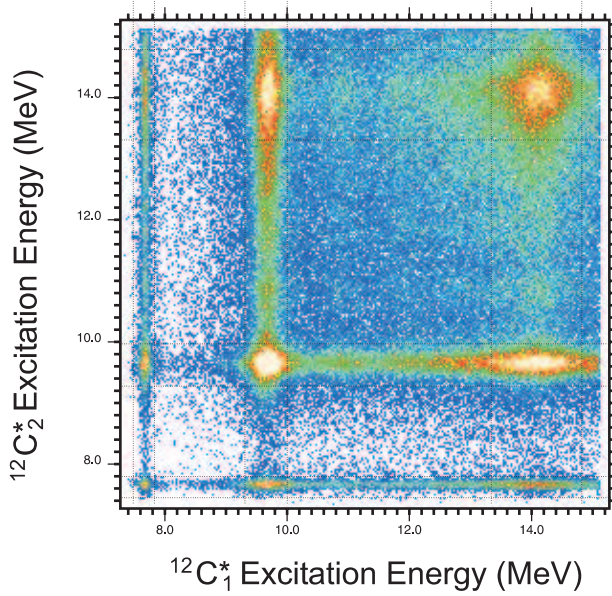
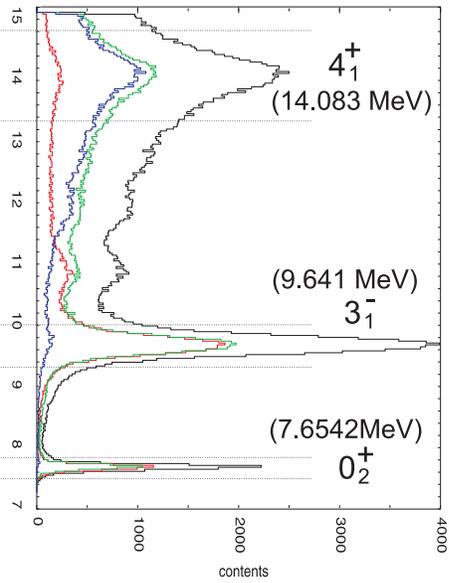
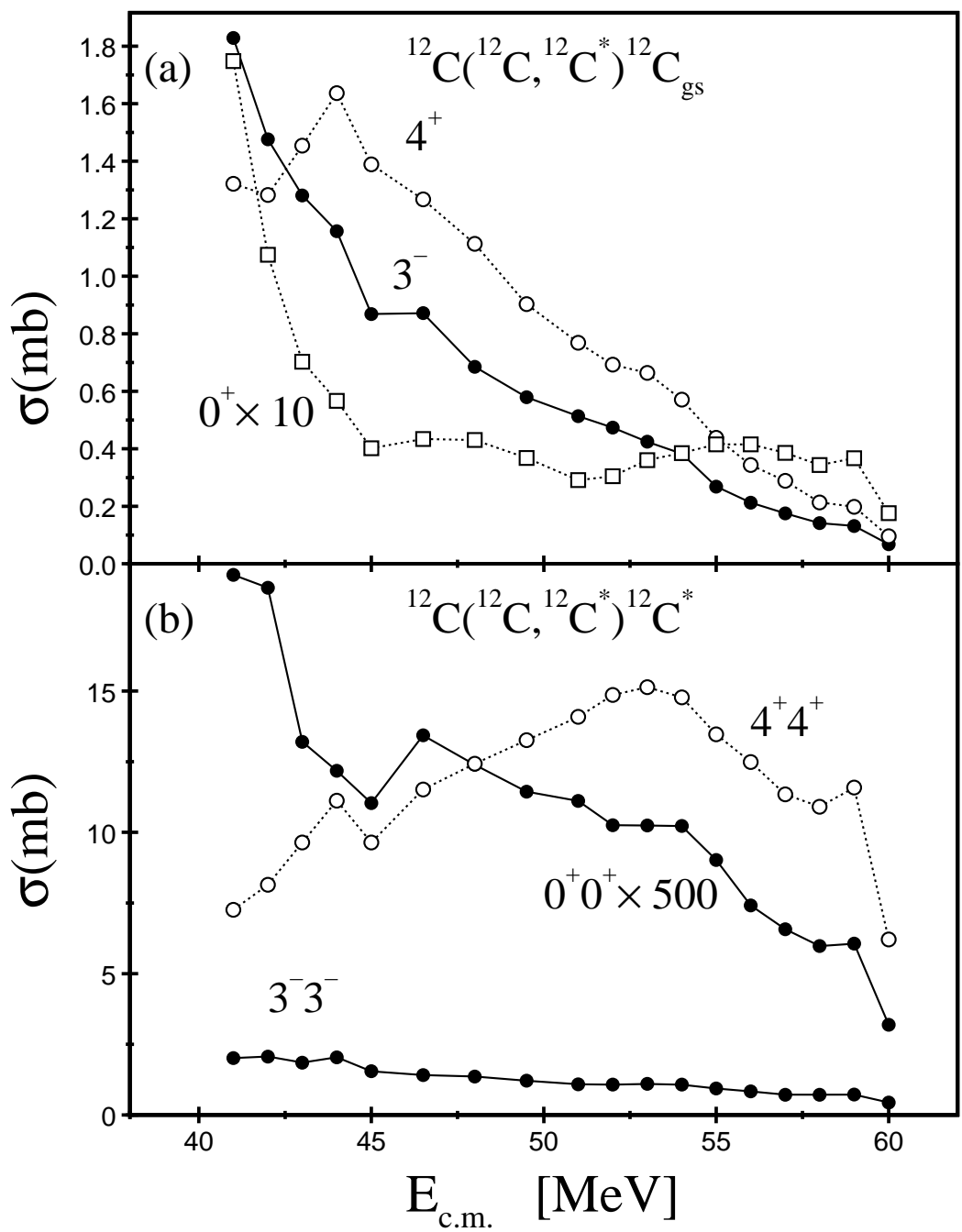
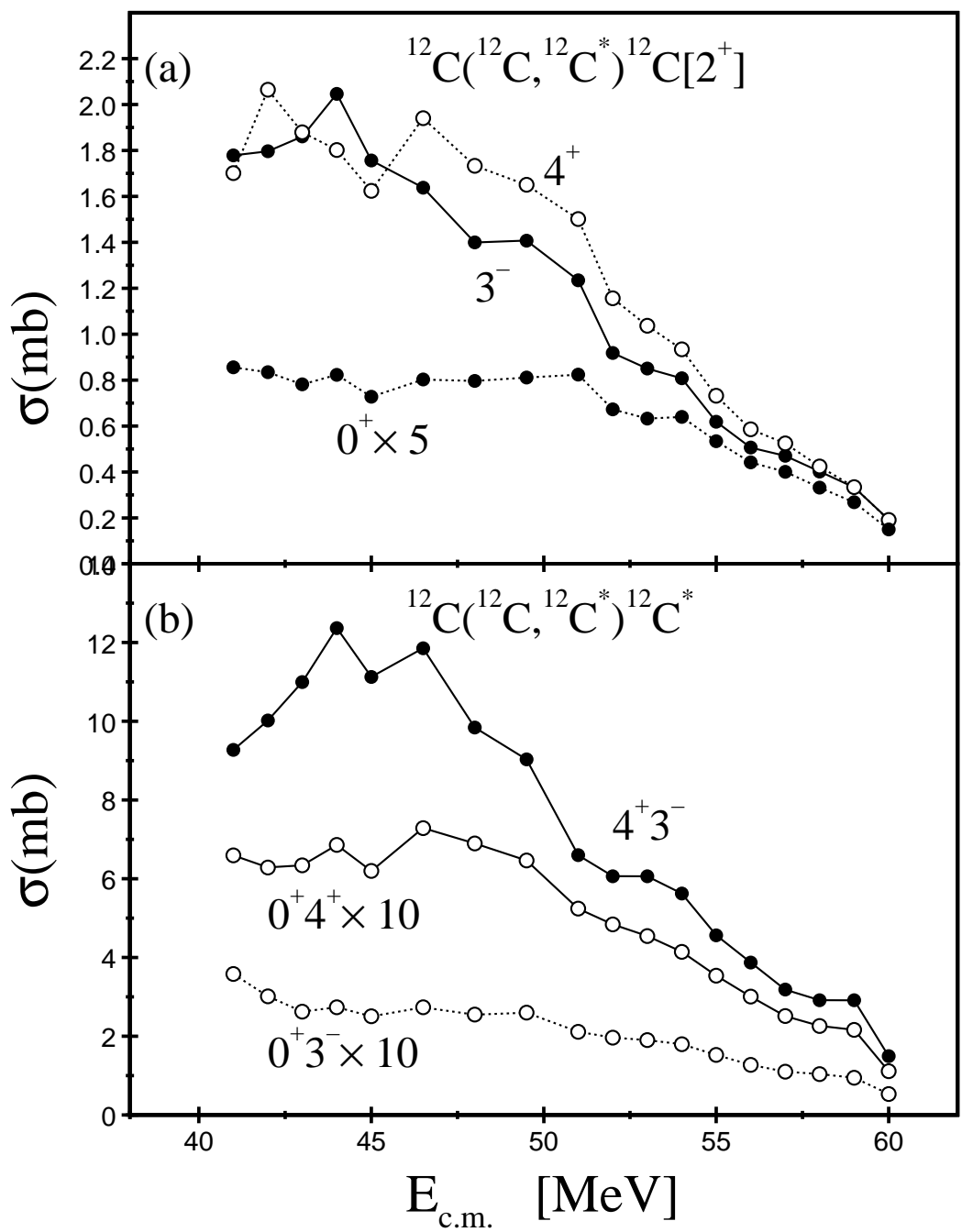
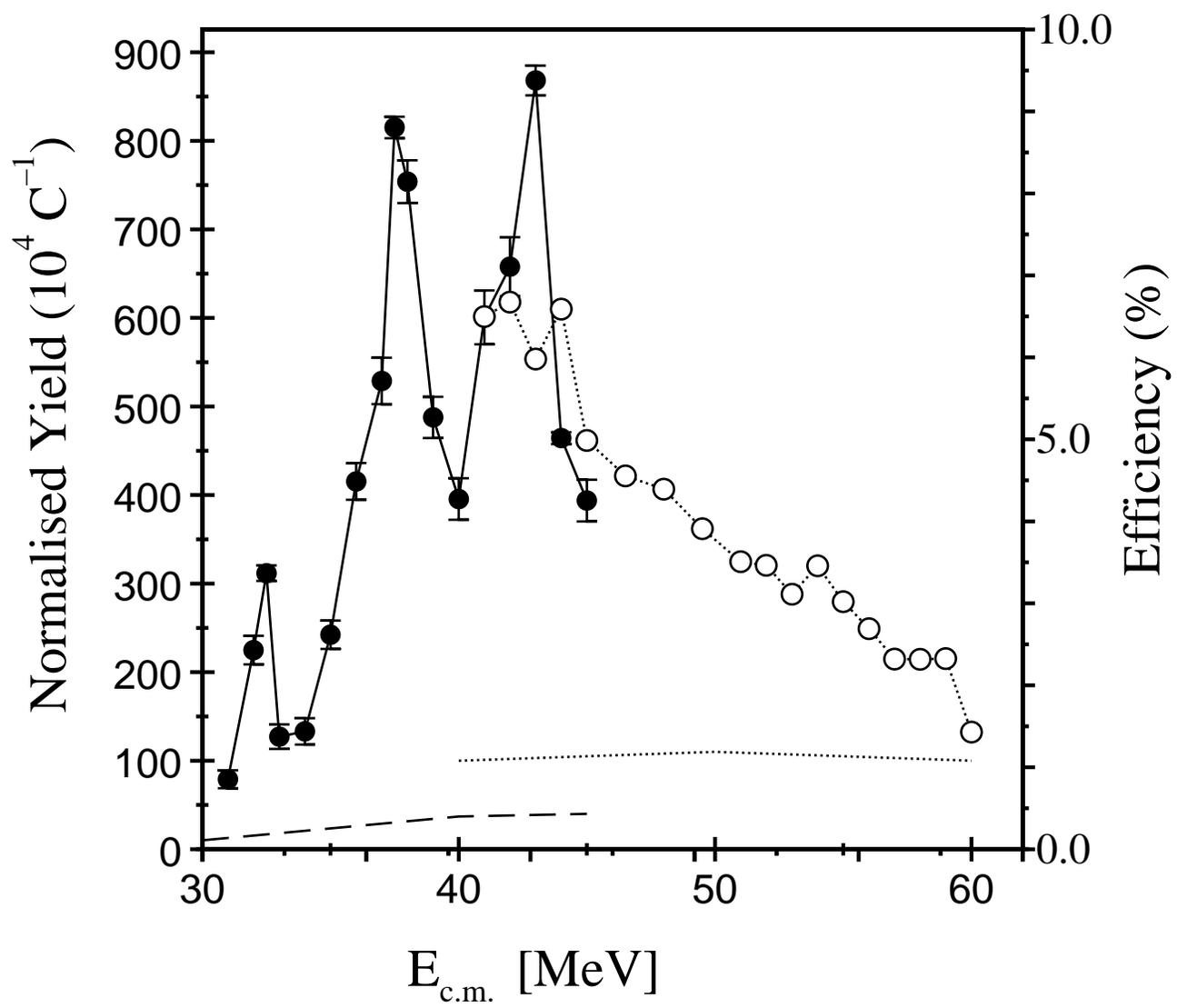
\end{document}